\documentclass[letter]{aa} 

\usepackage{graphicx}
\usepackage{txfonts}
\usepackage{natbib}

\bibpunct{(}{)}{;}{f}{}{,} 

\newcommand{\kms}{\,$\mathrm{km\, s^{-1}}$}
\newcommand{\ms}{\,$\mathrm{m\, s^{-1}}$}
\newcommand{\Teff}{\ensuremath{T_{\mathrm{eff}}}}
\newcommand{\logg}{\ensuremath{\log g}}

\newcommand{\Mj}{\ensuremath{M_{\mathrm{Jup}}}}

\usepackage{color}

\begin{document}

  \title{Search for giant planets in M67 III: \\
   excess of hot Jupiters in dense open clusters.
  \thanks{Based on observations collected at the ESO 3.6m telescope (La Silla), 
   at the 1.93m telescope of the Observatoire de Haute-Provence (OHP), 
   at the Hobby Eberly Telescope (HET), at the Telescopio Nazionale Galileo (TNG, La Palma) 
   and at the Euler Swiss Telescope.}}

  \author{A. Brucalassi\inst{1,2} \and  L. Pasquini\inst{3} \and R. Saglia\inst{1,2} \and M.T. Ruiz\inst{4} \and P. Bonifacio\inst{5}
  \and I. Le\~{a}o\inst{3,6}
  \and B.L. Canto Martins\inst{6} \and J.R. de Medeiros\inst{6} \and  L. R. Bedin\inst{7} \and K. Biazzo\inst{8} \and C. Melo \inst{9}\and 
  C. Lovis\inst{10} \and S. Randich\inst{11}}

   \institute{Max-Planck f\"ur extraterrestrische Physik, Garching bei M\"unchen, Germany
   \and University Observatory Munich, Ludwig Maximillian Universitaet, Scheinerstrasse 1, 81679 Munich, Germany
   \and ESO -- European Southern Observatory, Karl-Schwarzschild-Strasse 2, 85748 Garching bei M\"unchen, Germany 
   \and Astronomy Department, Universidad de Chile, Santiago, Chile 
   \and GEPI, Observatoire de Paris, CNRS, Univ. Paris Diderot, Place Jules Janssen 92190 Meudon, France
   \and Universidade Federal do Rio Grande do Norte, Natal, Brazil
   \and Istituto Nazionale di Astrofisica, Osservatorio Astronomico di Padova, Padova, Italy
   \and Istituto Nazionale di Astrofisica, Osservatorio Astronomico di Catania, Catania, Italy
   \and ESO -- European Southern Observatory, Santiago, Chile
   \and Observatoire de Geneve, Sauverny, CH
   \and Istituto Nazionale di Astrofisica, Osservatorio Astrofisico di Arcetri, Firenze, Italy
   }          
             
    \date{Received  / Accepted }


   \abstract{
   Since 2008 we used high-precision radial velocity (RV) measurements obtained with different telescopes to detect signatures of
   massive planets around main-sequence and evolved stars of the open cluster (OC) M67.
   We aimed to perform a long-term study on giant planet formation in open clusters 
   and determine how this formation depends on stellar mass and chemical composition.\\
   A new hot Jupiter (HJ) around the main-sequence star YBP401 is reported in this work.
   An update of the RV measurements for the two HJ host-stars YBP1194 and YBP1514 is also discussed.
   Our sample of 66 main-sequence and turnoff stars includes 3 HJs, 
   which indicates a high rate of HJs in this cluster (5.6$^{+5.4}_{-2.6}$\% for single stars and 4.5\%$^{+4.5}_{-2.5}$\% for the full sample ).
   This rate is much higher than what has been discovered in the field, either with RV surveys or by transits.
   High metallicity is not a cause for the excess of HJs in M67, nor can the excess be attributed to high stellar masses. 
   When combining this rate  with the non-zero eccentricity of the orbits, our results are 
   qualitatively consistent with a HJ formation scenario dominated by strong encounters with other 
   stars or binary companions and subsequent planet-planet scattering, as predicted by N-body simulations.
    }

  \keywords{Exoplanets -- Open clusters and associations: individual: M67 -- Stars: late-type -- Techniques: radial velocities}

   \maketitle
%

\section{Introduction}

 Hot Jupiters (HJs) are defined as giant planets ($M_{p} > 0.3$\Mj) on
 short-period orbits (P < 10 days).
 They show an occurrance rate of $\sim$1.2\% around Sun-like field stars \citep{Wright2012, Mayor2011}.
 These close-in giant planets are highly unlikely to have formed in situ,
 and it is believed that they form beyond the snow line 
 where solid ices are more abundant, allowing the planet cores
 to grow several times more massive than in the inner part of the proto-planetary disk
 before undergoing an inward migration.
 Of the mechanisms that are able to trigger migration,
 the two supported most often are dynamical interaction with the circumstellar disk \citep[\textit{\textup{type II migration}}]
 {Goldreich80,Lin86,Ward97}
 and gravitational scattering caused by other planets \citep[\textit{\textup{planet-planet scattering}}]{Rasio96,Lin97}.
 Other ideas include violent migration mechanism such as dynamical encounters
 with a third body (multi-body dynamical interaction).
 In particular, recent N-body simulations have shown that a planetary system inside a crowded birth-environment 
 can be strongly destabilized by stellar encounters and dynamical interaction, 
 which also favours the formation of HJs \citep{Davies2014, Malmberg2011,Shara2016}.
 Open clusters (OCs) hold great promise as laboratories in which properties of exoplanets and
 theories of planet formation and migration can be explored.
 
  In Paper I \citep{Pasquini2012} we described
  a radial velocity (RV) survey to detect the signature of giant planets around
  a sample of main-sequence (MS) and giant stars in M67.
  The first three planets discovered were presented in \citet{Brucalassi2014}.
  One goal of this project is to investigate whether and how  planet formation is influenced by the environment.
  Recent planet search surveys in OCs support that 
  the statistics in OCs is compatible with the field \citep{Malavolta2016,Brucalassi2014,Meibom2013,Quinn2014,Quinn2012}. 
  In this work we show that for M67 the frequency of HJs is even higher than in the field. 
   
\section{Observations and orbital solutions}   
\label{sec:Sample_Obs}

  Of the 88 stars in the original M67 sample,
  12 have been found to be binaries \citep{Pasquini2012}.
  Two additional binaries have recently been discovered (Brucalassi et al. 2016).
  The final sample therefore comprises 74 single stars (53 MS and turnoff stars and 21 giants)
  that are all high-probability members (from proper motion and radial velocity) of the cluster
  according to \citet{Yadav2008} and \citet{Sanders77}.

 \begin{table}
\caption{Stellar parameters of the three M67 stars newly found to host planet candidates.}
\label{StarParam}
\centering
\small
\begin{tabular}{lrrr}
\hline
\textbf{Parameters}&YBP401&YBP1194&YBP1514\\
\hline
$\mathrm{\alpha}$ $(\mathrm{J2000})$&  08:51:19.05 &   08:51:00.81   &  08:51:00.77  \\
$\mathrm{\delta}$ $(\mathrm{J2000})$& +11:40:15.80 &  +11:48:52.76   & +11:53:11.51  \\
Spec.type       &F9V    &G5V& G5V\\
\ensuremath{m_{\mathrm{V}}}  $[\mathrm{mag}]$          &13.70\tablefootmark{a}            &14.6\tablefootmark{a}           &14.77\tablefootmark{a}        \\
\ensuremath{B-V}             $[\mathrm{mag}]$          &0.607\tablefootmark{a}            &0.626\tablefootmark{a}          &0.680\tablefootmark{a}        \\
\ensuremath{M\star}           [\ensuremath{M_{\odot}}] &1.14$\pm$0.02\tablefootmark{b}    &1.01$\pm$0.02\tablefootmark{b}  &0.96$\pm$0.01\tablefootmark{b}\\
$\logg$                      $[\mathrm{cgs}]$          &4.30$\pm$0.035\tablefootmark{d}   &4.44$\pm$0.05\tablefootmark{c}  &4.57$\pm$0.06\tablefootmark{d}\\
$\Teff$                      $[\mathrm{K}]$            &6165$\pm$64\tablefootmark{d}      &5780$\pm$27\tablefootmark{c}    &5725$\pm$45\tablefootmark{d}  \\
\hline
\end{tabular}                                                                                                         
\normalsize 
\tablefoot{\tablefoottext{a}{\citet{Yadav2008}.}
\tablefoottext{b}{\citet{Pietrinferni2004} and \citet{Girardi2000}.}
\tablefoottext{c}{\citet{Onehag2011}.}
\tablefoottext{d}{\citet{Pasquini2008} and \citet{Pace2012}.}
}
\end{table}
 The star YBP401
 shows significant indications of a HJ companion
 and is analysed here in detail.
 We also present an update of the RV measurements for the stars YBP1194 and YBP1514, 
 for which two other HJs were announced in our previous work \citep{Brucalassi2014}.
 
 Basic stellar parameters 
 ($V$, $B-V$, $\Teff$, $\logg$) 
 with their uncertainties were taken from
 the literature. A distance modulus of 9.63$\pm$0.05 \citep{Pasquini2008} and
 a reddening of E(B-V)=0.041$\pm$0.004 \citep{Taylor2007} were assumed,
 stellar masses and radii were derived 
 using the 4 Gyr theoretical isochrones from \citet{Pietrinferni2004} and \citet{Girardi2000}.
 The parameters estimated from isochrone fitting agree within the errors with the values
 adopted from the literature. 
 The main characteristics of the three host stars are listed in Table~\ref{StarParam}.

 \begin{figure}
 \vspace{-5pt}
 \centering
  \hspace{-16pt}
 \includegraphics[width=0.48\textwidth,angle=0]{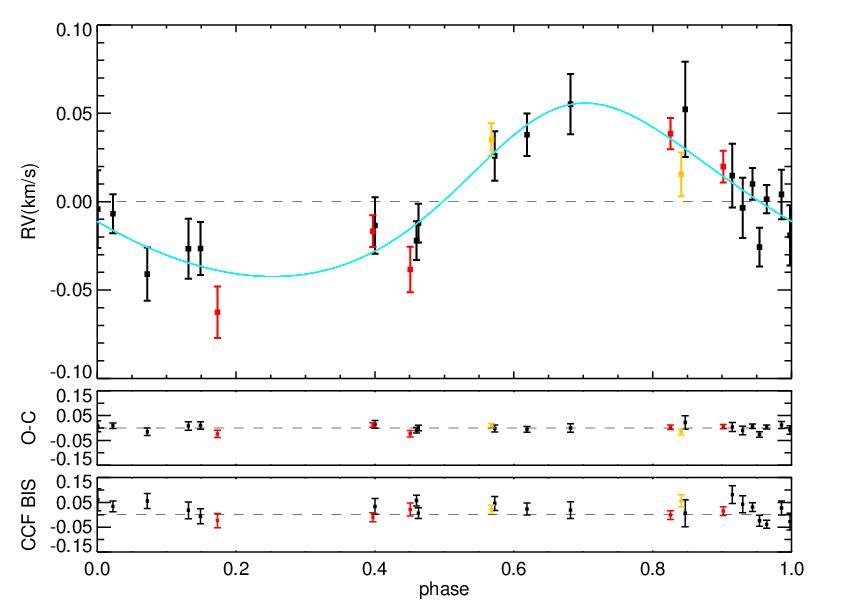}
 \vspace{-5pt}
 \caption{ 
  Phased RV measurements and Keplerian best fit, best-fit residuals, and
  bisector variation for YBP401. Black dots: HARPS measurements, red
  dots: SOPHIE measurements, orange dots: HARPS-N measurements. }
  \vspace{-5pt} 
 \label{Fit_YBP401}
 \end{figure}
 The RV measurements were carried out using the HARPS spectrograph \citep{Mayor03} at the ESO 3.6m telescope
 in high-efficiency mode (with R=90\,000 and a spectral range of 378-691 nm), with 
 the SOPHIE spectrograph \citep{Bouchy06} at the OHP 1.93 m telescope in high-efficiency mode
 (with R=40\,000 and a range of 387-694 nm),
 with the HRS spectrograph \citep{Tull1998} at the Hobby Eberly Telescope
 (with R=60\,000 and a range of 407.6-787.5 nm),
 and with the HARPS-N spectrograph at the TNG on La Palma of
 the Canary Islands
 (spectral range of 383-693 nm and R=115\,000).
 Additional RV data points for giant stars
 have been observed between 2003 and 2005 \citep{Lovis2007}
 with the CORALIE spectrograph at the 1.2 m Euler Swiss telescope.
 
 HARPS, SOPHIE, and HARPS-N are provided with a similar automatic pipeline. 
 The spectra are extracted from the detector images and
 cross-correlated with a numerical G2-type mask. 
 Radial velocities are derived by fitting
 each resulting cross-correlation function (CCF) with a Gaussian
 \citep{Baranne1996,Pepe2002}.
 For the HRS, the radial velocities were computed 
 using a series of dedicated routines based on IRAF and by cross-correlating
 the spectra with a G2 star template \citep{Cappetta2012}.
 We used nightly observations of the RV standard star HD32923
 to correct all observations for each
 star to the zero point of HARPS \citep[as explained
 in][]{Pasquini2012} and to take into account any 
 instrument instability or systematic velocity shifts between runs.
 An additional correction was applied to the SOPHIE data to
 consider the low signal-to-noise ratio (S/N) of the
 observations \citep[see Eq.(1)]{Santerne2012}.

   \begin{figure}
   \vspace{-5pt}
   \centering
    \hspace{-16pt}
    \includegraphics[width=0.48\textwidth]{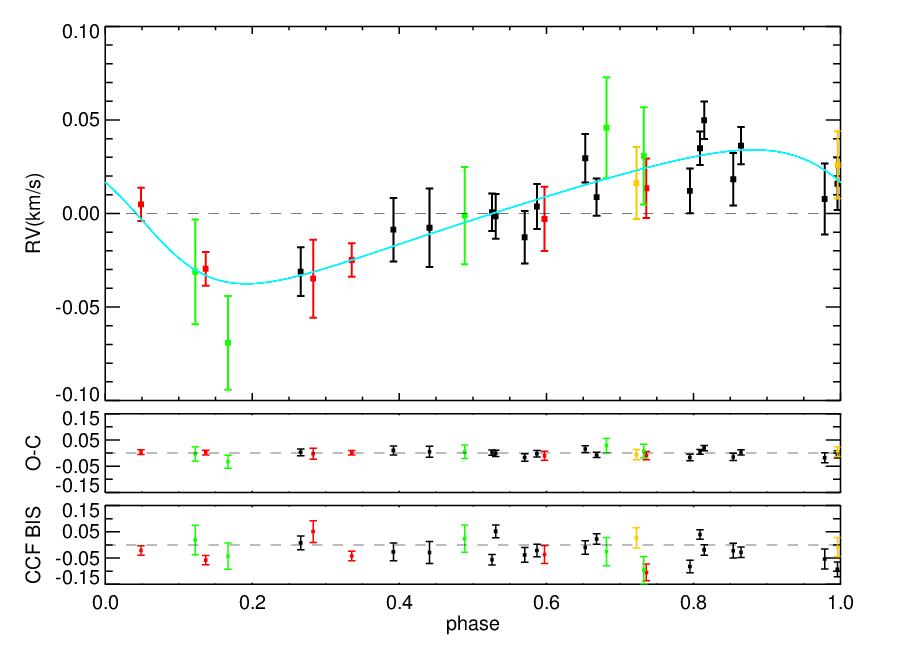}
      \vspace{-5pt}
      \caption{Phased RV measurements and Keplerian best fit, best-fit residuals, and 
      bisector variation for YBP1194.
      Same symbols as in Fig.\ref{Fit_YBP401}, green dots: HRS measurements.}
      \vspace{-5pt} 
         \label{FIT_YBP1194}
   \end{figure}
\begin{figure*}[ht]
                \begin{center}
                        \begin{tabular}{c}
                               \hspace{.4cm}
                                RV values \hspace{3.0cm} Residuals \hspace{3.0cm} BIS span \hspace{3.0cm} FWHM\\

 \resizebox{\hsize}{!}
            {
    \includegraphics{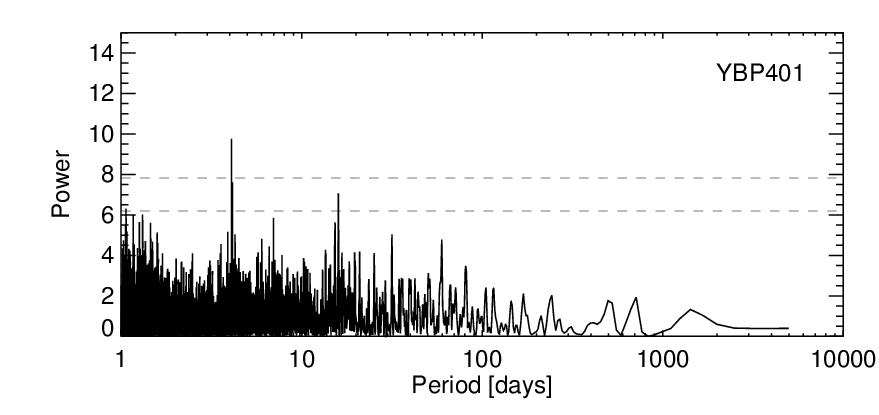}
    \includegraphics{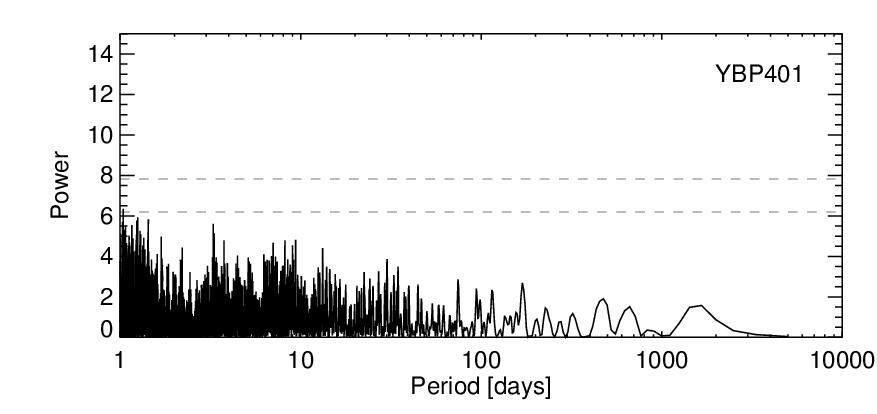}
    \includegraphics{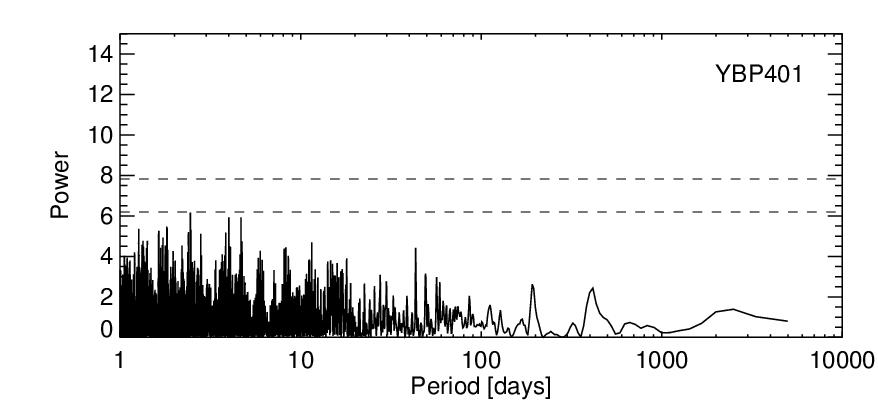} 
    \includegraphics{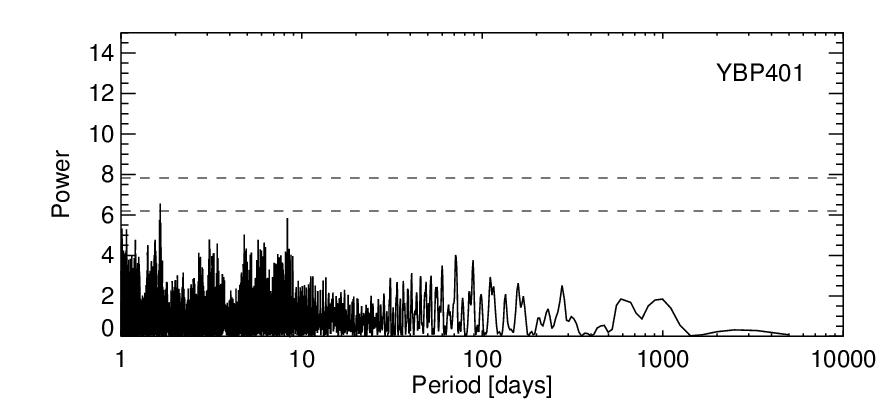} 
  }\\ 
  \resizebox{\hsize}{!}
       {
   \includegraphics{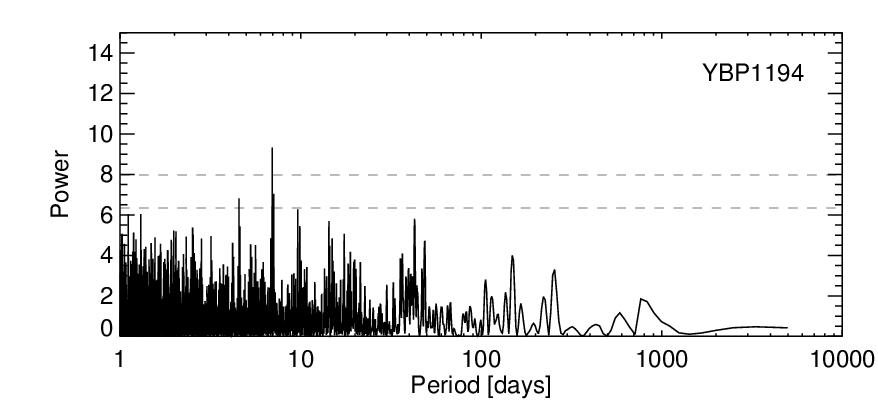}
    \includegraphics{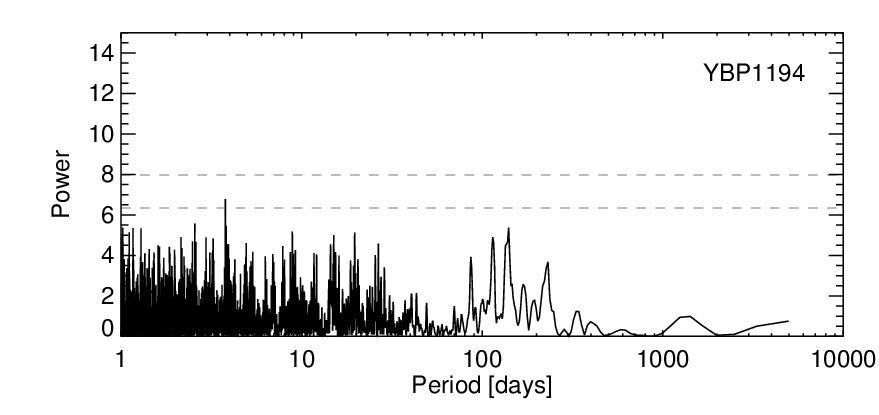}
    \includegraphics{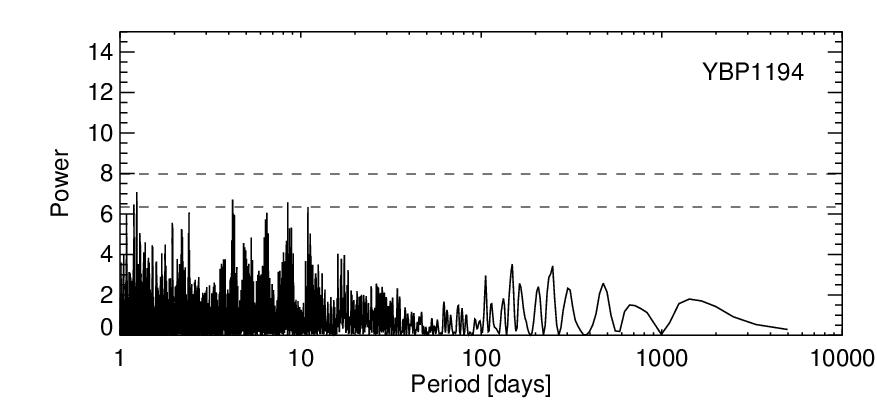} 
    \includegraphics{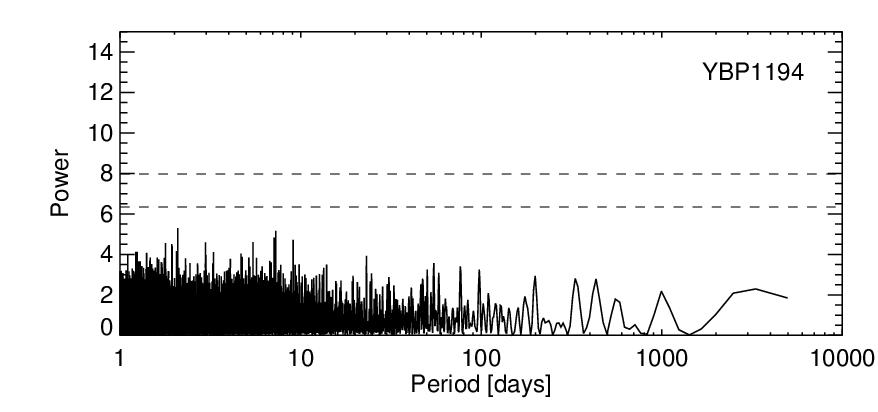}  
       }\\
    \resizebox{\hsize}{!}
       {
   \includegraphics{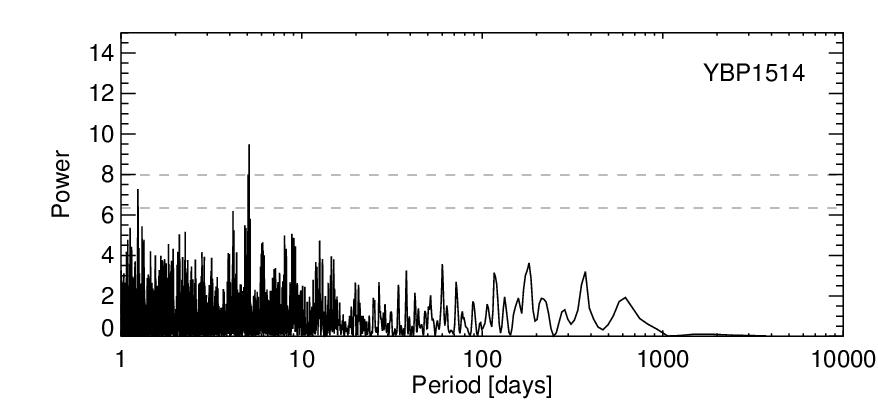} 
    \includegraphics{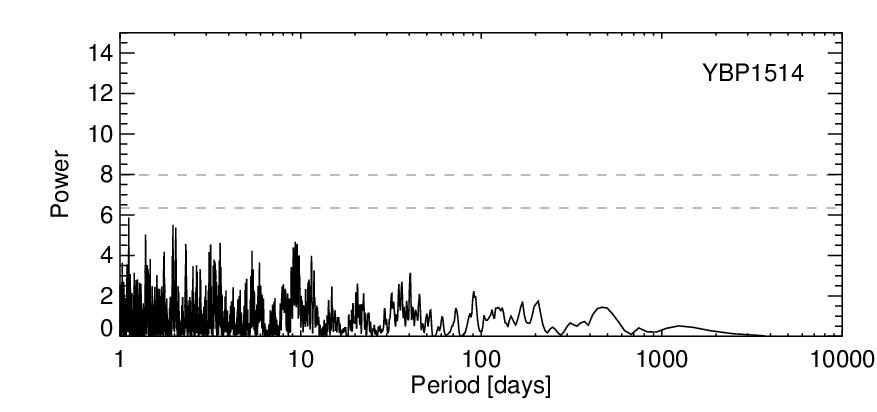} 
    \includegraphics{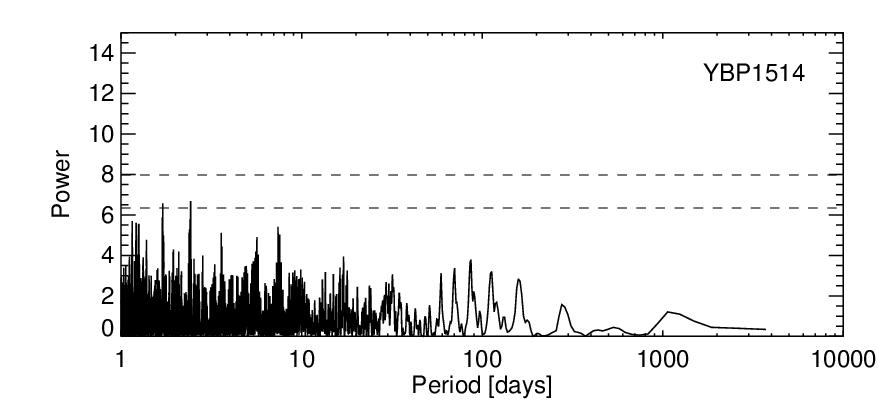}  
    \includegraphics{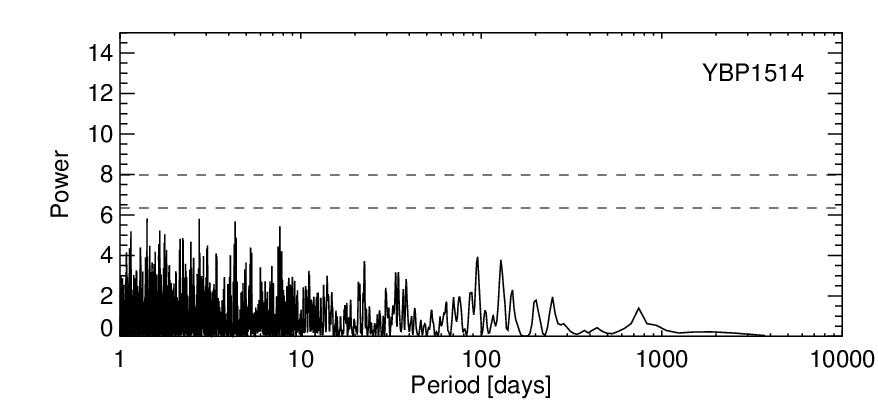}   
       } \\ 

                        \end{tabular}
                \end{center}
                \caption
           { \label{Periodograms} 
           Top: Lomb-Scargle periodogram of the RV measurements, residuals, bisector span, and FWHM for YBP401.
     Central: Same plots for YBP1194. Bottom: same plots for YBP1514. 
     The dashed lines correspond to 5\% and 1\% false-alarm probabilities, calculated according 
     to Horne \& Baliunas (1986) and white noise simulations.
                }
        \end{figure*}
\begin{figure*}[h]
 \centering
 \resizebox{\hsize}{!}
            {
 \includegraphics{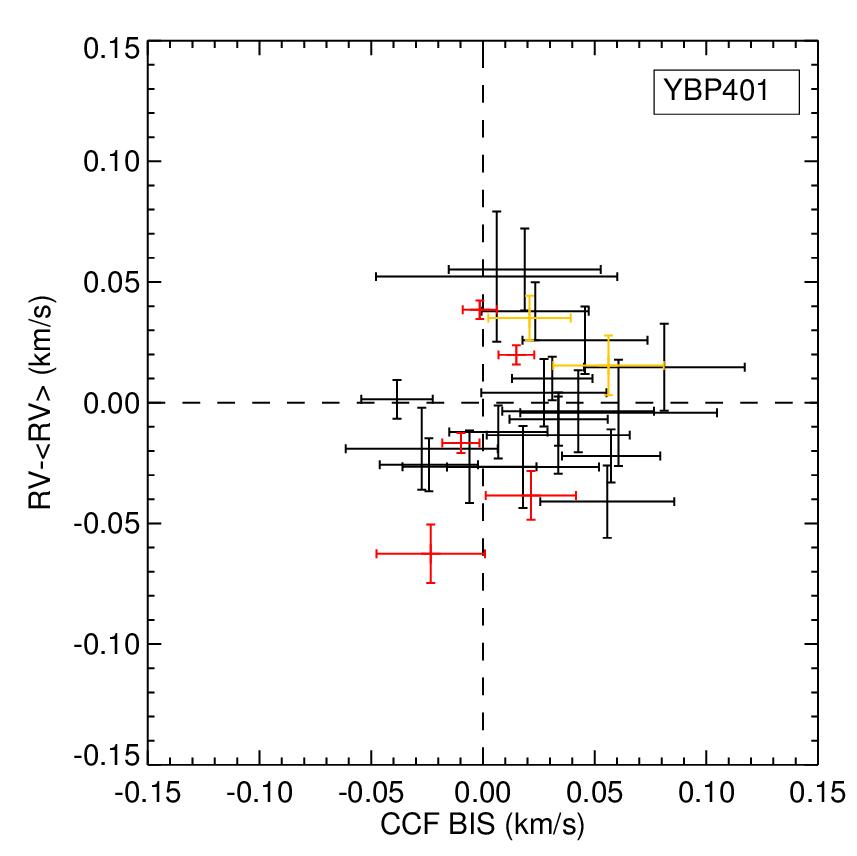}
 \includegraphics{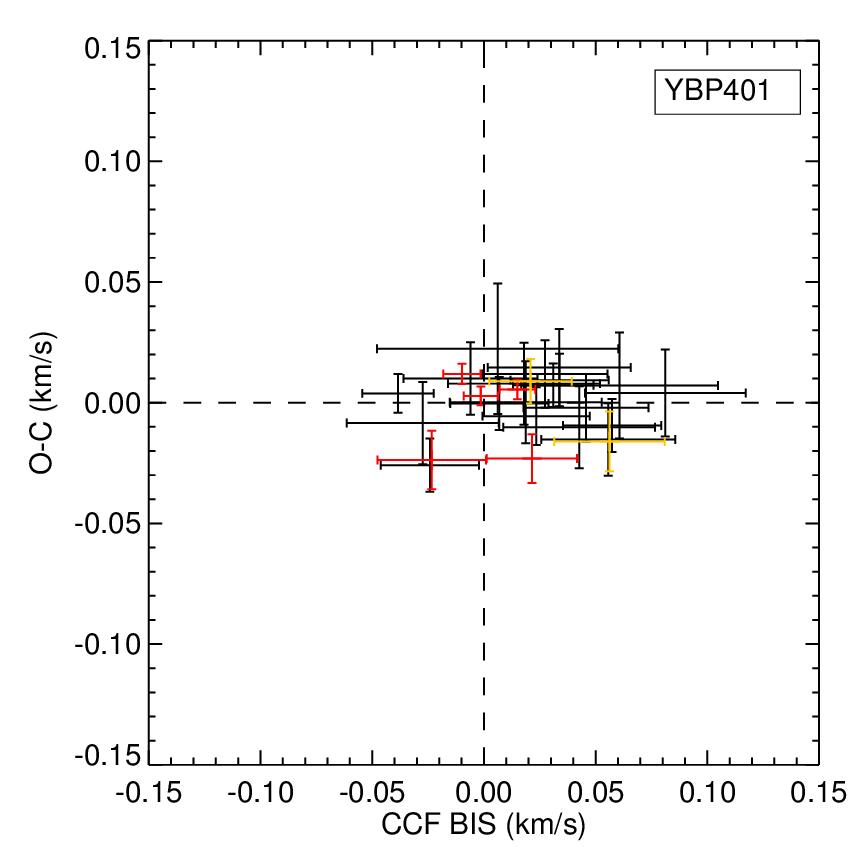}
 \includegraphics{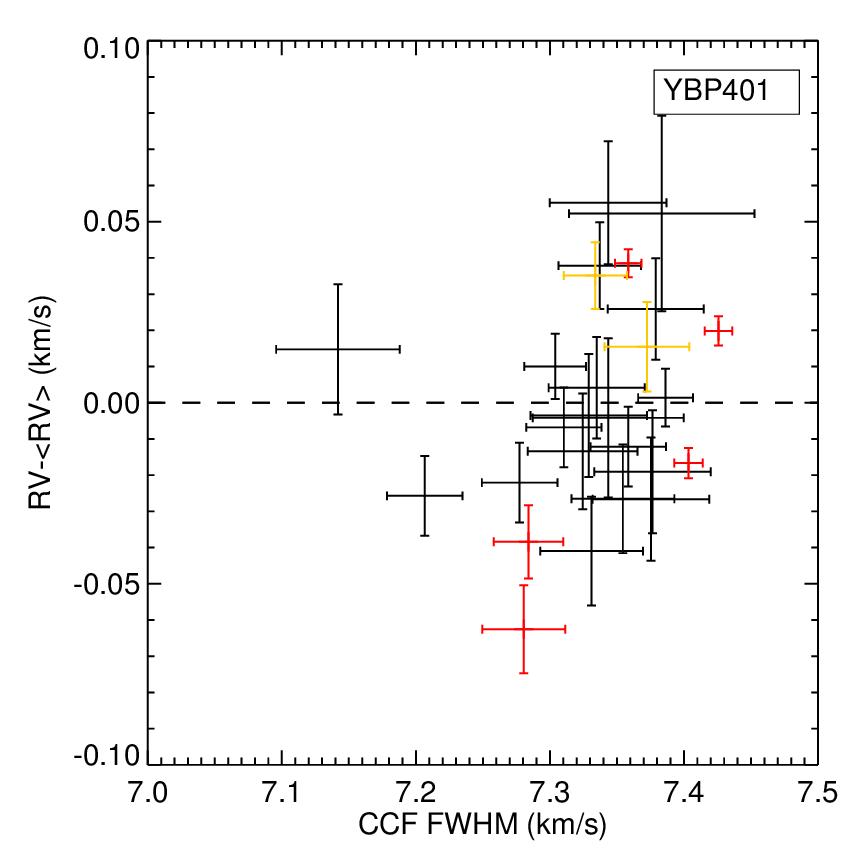}
 \includegraphics{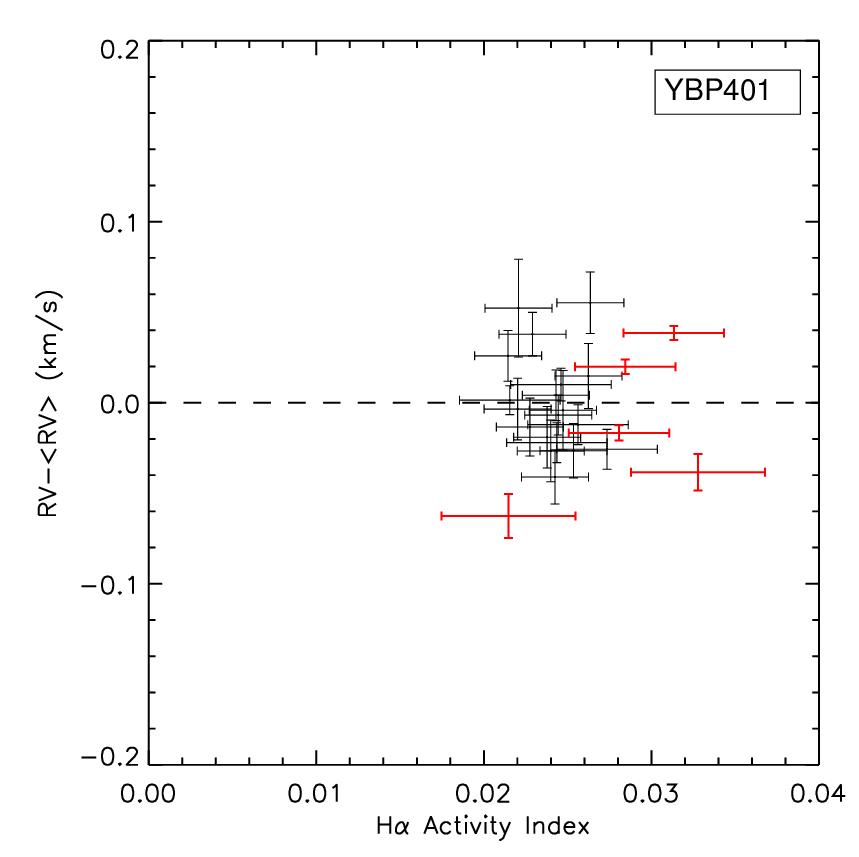}
  }
 \resizebox{\hsize}{!}
            {
 \includegraphics{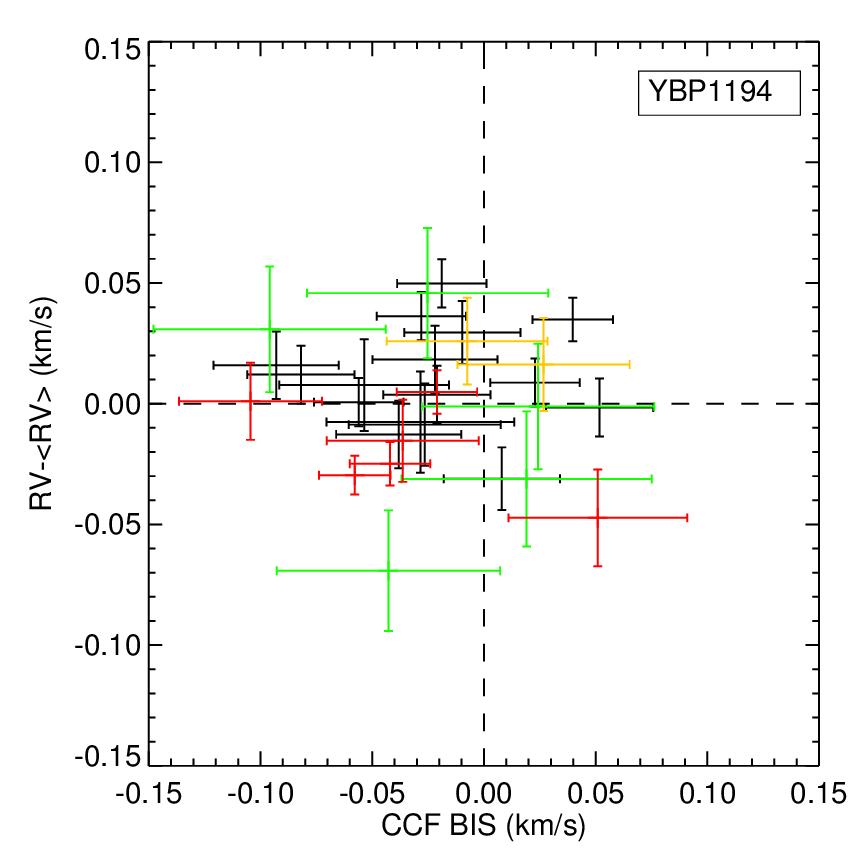}
 \includegraphics{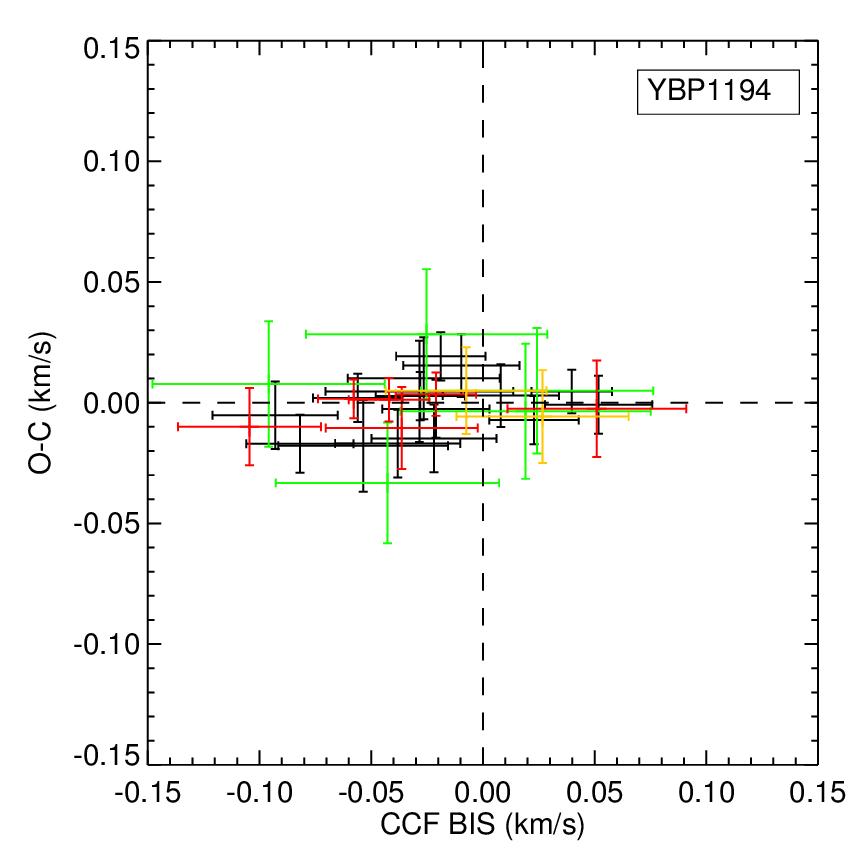}
 \includegraphics{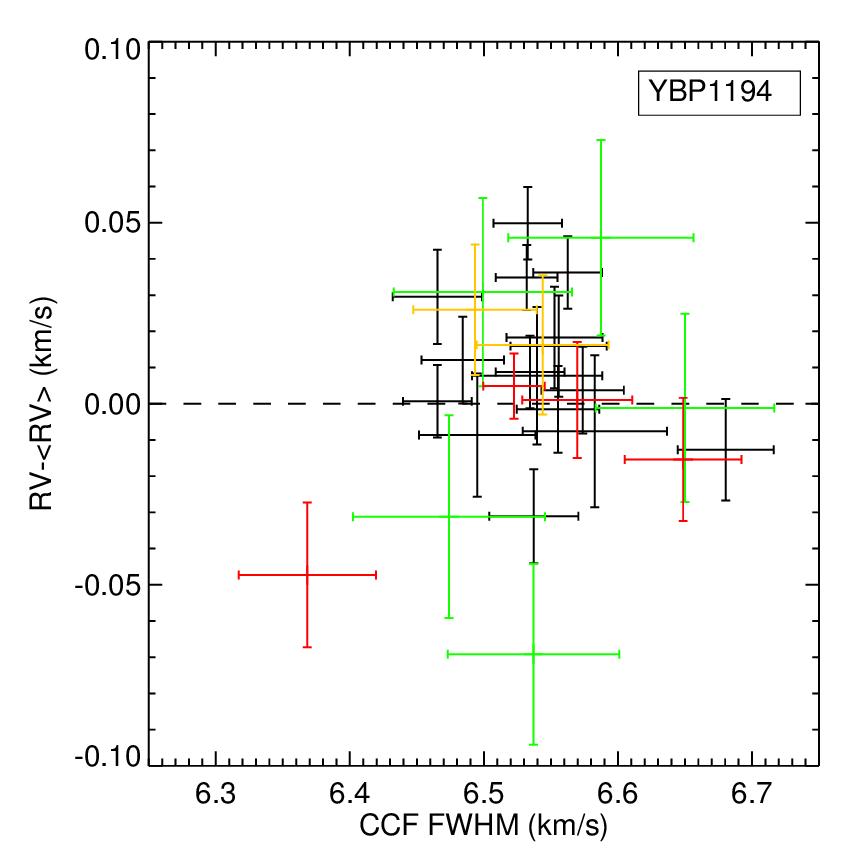}
 \includegraphics{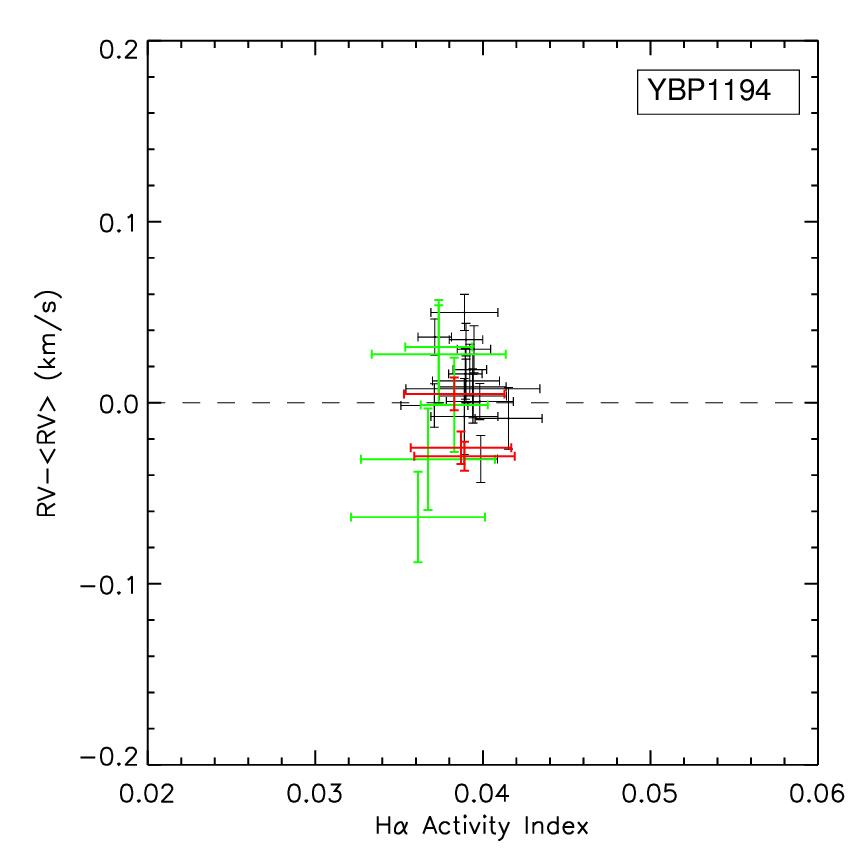}
  }
 \resizebox{\hsize}{!}
            {
 \includegraphics{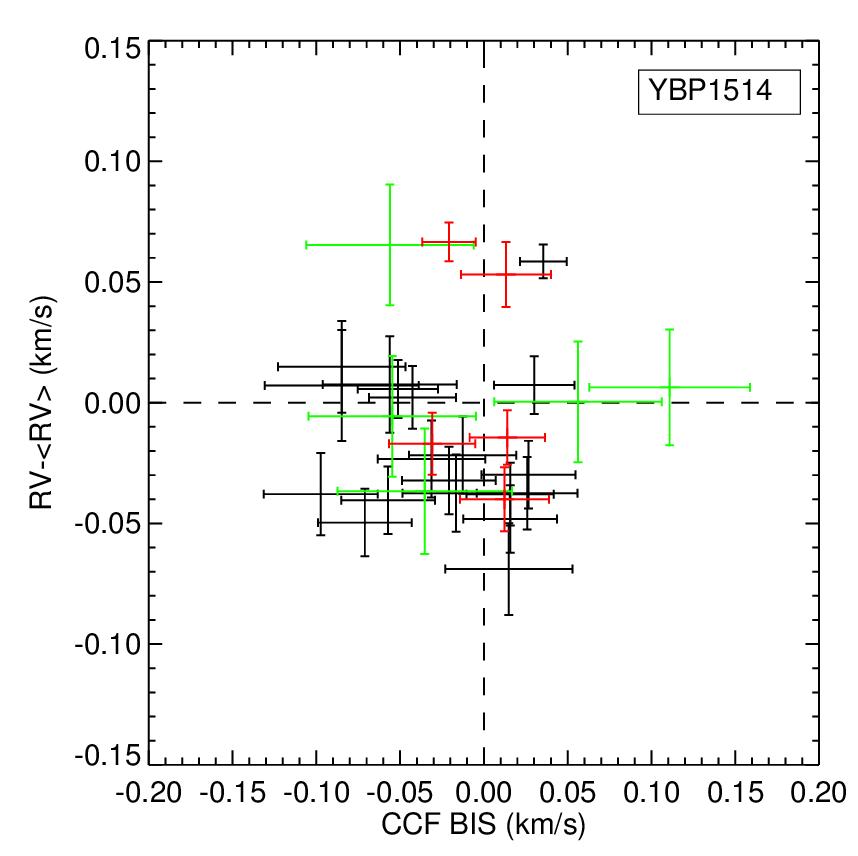}
 \includegraphics{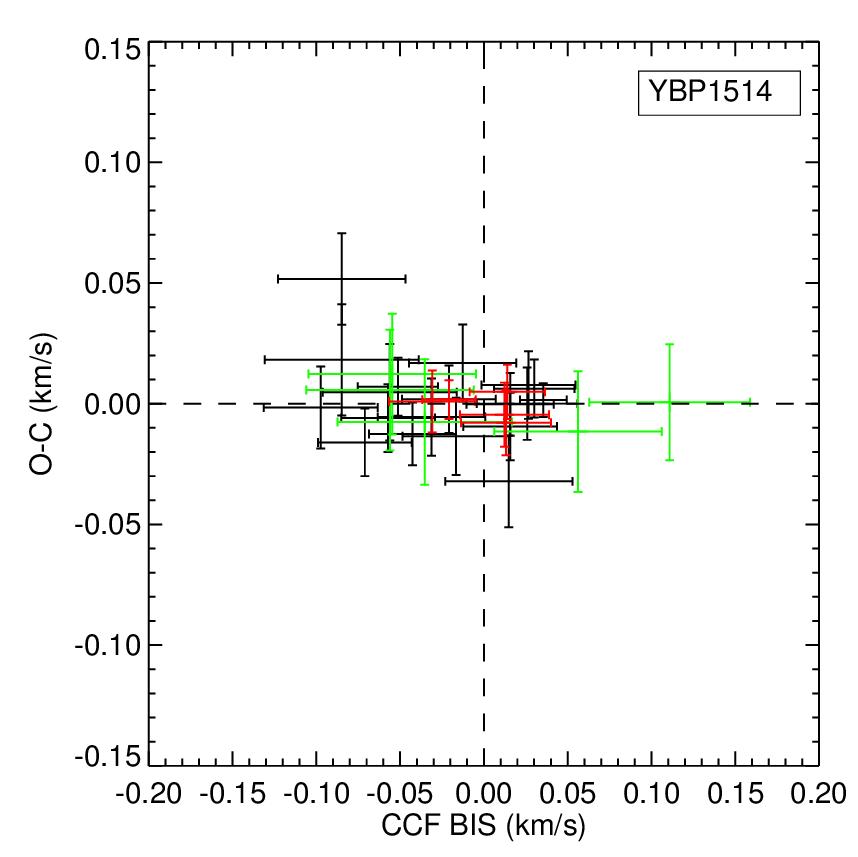}
 \includegraphics{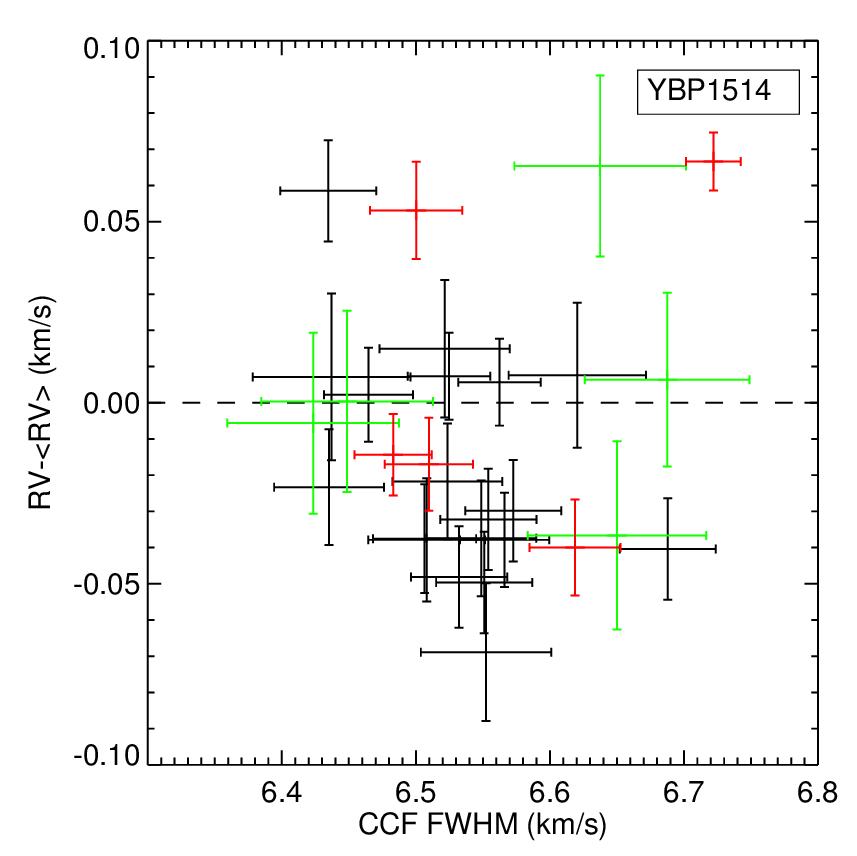}
 \includegraphics{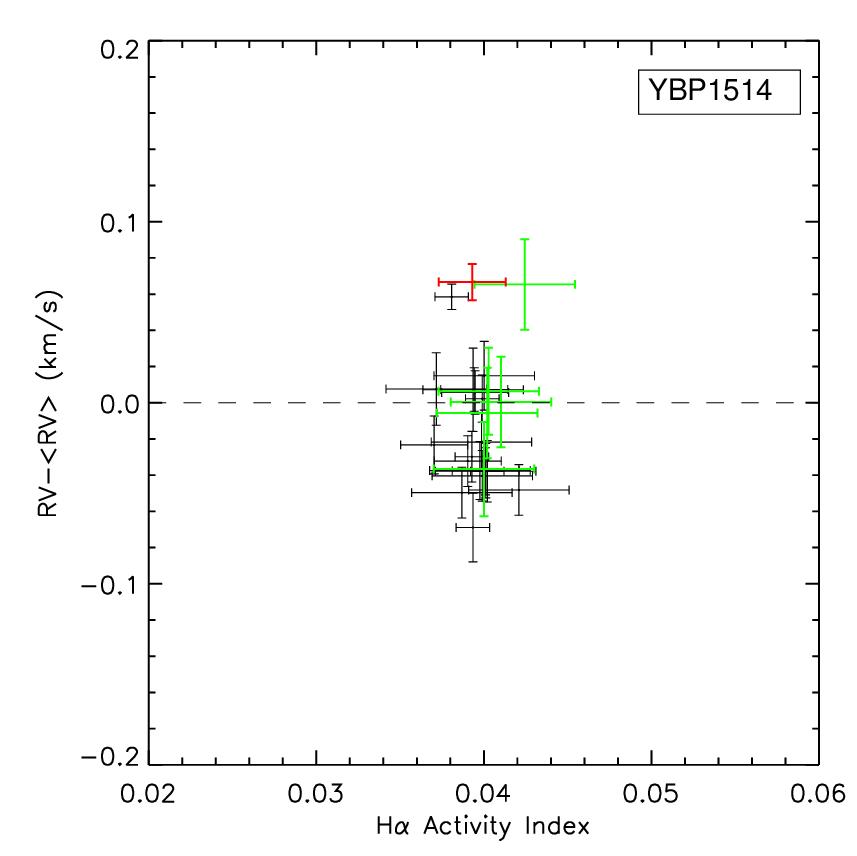}
  } 
  
 \caption{Top: RV measurements versus bisector span, residuals versus bisector span, RV measurements versus CCF FWHM and 
 RV measurements versus H$\alpha$ activity indicator for YBP401. The H$\alpha$ activity indicator is computed as the area
 below the core of H$\alpha$ line with respect to the continuum. 
 CCF FWHM values are calculated by subtracting the respective instrumental FWHM in quadrature.
 Same symbols as in Fig.~\ref{FIT_YBP1194}. 
 Middle: The same plots for YBP1194. Bottom: The same plots for YBP1514.
 }
 \label{BIS_FWHM_Ha}
 \end{figure*}

%
%
%
 The RV measurements of our target stars were studied  
 by computing the Lomb-Scargle periodogram \citep{Scargle1982,Horne1986} and by using a
 Levenberg-Marquardt analysis \citep[RVLIN]{Wright2009} to fit Keplerian orbits to the radial velocity data.
 The orbital solutions were independently checked using the Yorbit
 program (Segransan et al. 2011) and a simple Markov chain Monte Carlo (MCMC) analysis
 (see Table~\ref{PlanetParamMCMC}).
 We investigated the presence and variability of
 chromospheric active regions  in  these  stars
 by measuring the variations of the core of the H$\alpha$ line
 with respect to the continuum,
 following a method similar to the one described 
 in \citet{Pasquini1991}.  
 The more sensitive Ca II H and K lines were not accessible
 because of the low S/N ratio of our observations.
 For each case we verified the correlation between the RVs and the bisector span 
 of the CCF \citep[calculated following][]{Queloz01} or 
 with the full width at half maximum (FWHM) of the CCF.\\
 \\
 \textbf{YBP401.} According to \citet{Yadav2008}, this F9V MS star has 
 a membership probability of 97\% and
 a proper motion shorter than 6 mas/yr with respect to the average.
 \citet{Vereshchagin2014} revised the membership list of \citet{Yadav2008}
 and expressed doubts about the cluster membership for YBP401.
 However, the RV value considered for YBP401 in \citet{Vereshchagin2014}
 has an uncertainty of $\sigma=130$\ms\, and
 is not consistent with our measurements by more than 1$\sigma$.
 Recently, \citet{Geller2015} confirmed YBP401 as a single cluster member.\\
 This target has been observed since January 2008:
 19 RV points have been obtained with HARPS with a typical S/N of 15 (per
 pixel at 550 nm) and a mean measurement uncertainty
 of 15\ms\, including calibration errors. 
 Five additional RV measurements were obtained with SOPHIE and two with HARPS-N, with measurement uncertainties of 9.0\ms\, and 11.0\ms\, , respectively.
%
   \begin{figure}
   \vspace{-5pt}
   \centering
   \hspace{-16pt}
   \includegraphics[width=0.48\textwidth]{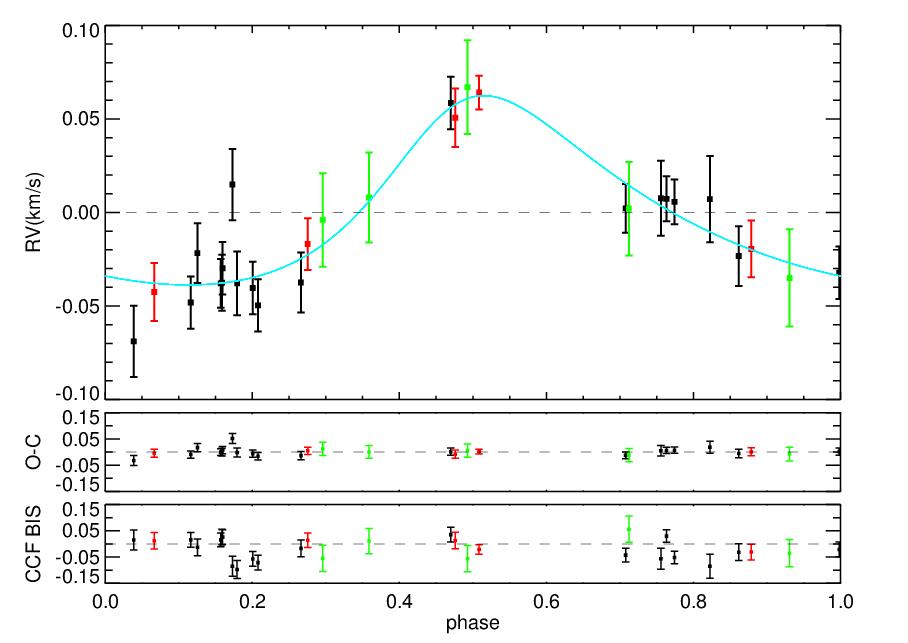}
    \vspace{-5pt}
      \caption{Phased RV measurements and Keplerian best fit, best-fit residuals, and 
       bisector variation for YBP1514. Same symbols as in Fig.\ref{FIT_YBP1194}.
               }
     \vspace{-5pt}          
         \label{FIT_YBP1514}
   \end{figure}
%
%
 The final 26 RV measurements of YBP401 show a variability of $\sim$35\ms and
 an average uncertainty of $\sim$14\ms\, for the individual RV values.
 A clear peak is present in the periodogram (see Fig.~\ref{Periodograms})
 at 4.08 days. 
 A Keplerian orbit was adjusted to the RV data of YBP401 (see Fig.~\ref{Fit_YBP401}), and
 the resulting orbital parameters for the planet candidate are reported in Tables~\ref{PlanetParam} and~\ref{PlanetParamMCMC}.
 We  note that the non-zero eccentricity is consistent with e$=$0 within 2$\sigma$
 and the other parameters change by less than  1$\sigma$ when fixing e$=$0.
 We included the eccentricity in the data analysis, which resulted in a better fit ($\chi_{red}^{2}\sim1$) 
 and in reduced RV residuals. However, more precise observations are needed to constrain 
 small non-zero eccentricities and to avoid overinterpreting the results
 (see discussions in Zakamska et al. 2011; Pont et al. 2011).
 The residuals have an rms amplitude of $\sim$13\ms\, and
 the periodogram of the residuals does not show any clear periodicity when the main signal is removed (see Fig.~\ref{Periodograms}).
 Neither the bisector span nor the activity index present correlations with the RV variations (see Fig.~\ref{BIS_FWHM_Ha}); this 
 excludes activity-induced variations of the shape or the spectral lines as the source of the RV measurements.\\
  \\
 \textbf{YBP1194 and YBP1514.} We have now collected 29 measurements 
 for both YBP1194 and YBP1514, spanning seven years.
 The average RV uncertainty is $\sim$13.0\ms\, for HARPS and SOPHIE,
 $\sim$26.0\ms\, for HRS and $\sim$8.0\ms\, for HARPS-N .
 Figures~\ref{FIT_YBP1194} and \ref{FIT_YBP1514} 
 show the phase-folded data points together with the best-fit solution 
 and the residual over the time.
 The peak in the periodogram is more pronounced and the RV
 signal is better determined (see Fig.~\ref{Periodograms}) than in \citet{Brucalassi2014}. 
 When the planet signature is removed,
 the rms of the residuals is $\sim$12.3\ms\, for YBP1194 and $\sim$14.4\ms\, for YBP1514.\\
 We note that the resulting updated orbital parameters are consistent within the errors with
 the previously published data (see Tables~\ref{PlanetParam} and~\ref{PlanetParamMCMC}).
 \begin{table}
\caption{Orbital parameters of the planetary candidates. \ensuremath{P}: period, \ensuremath{T}: time at periastron passage,
 \ensuremath{e}: eccentricity, $\omega$: argument of periastron, \ensuremath{K}: semi-amplitude of the RV curve, 
 \ensuremath{m\sin{i}}: planetary minimum mass,
 $\gamma$: average radial velocity, $\sigma$(O-C): dispersion of Keplerian fit residuals.  }
\label{PlanetParam}
\centering
\resizebox{0.5\textwidth}{!}{%
\vspace{-5pt}
\begin{tabular}{lrrr}
\hline
\textbf{Parameters} &YBP401&YBP1194&YBP1514\\
\hline
\ensuremath{P}        $[\mathrm{days}]$& 4.087$\pm$0.007   & 6.959$\pm$0.001     & 5.118$\pm$0.001   \\
\ensuremath{T}        $[\mathrm{JD}]$  & 2455974.3$\pm$0.5 & 2455289.98$\pm$0.51 & 2455986.3$\pm$0.3 \\
\ensuremath{e}                         & 0.15$\pm$0.08     & 0.30$\pm$0.08       & 0.28$\pm$0.09     \\
$\omega$              $[\mathrm{deg}]$ & 330.17$\pm$25.68  & 99.36$\pm$20.47     & 327.33$\pm$16.09  \\
\ensuremath{K}        [\ms]            & 49.06$\pm$3.50    & 35.80$\pm$3.81      & 50.06$\pm$5.03   \\
\ensuremath{m\sin{i}} [\Mj]            & 0.46$\pm$0.05     & 0.32$\pm$0.03       & 0.40$\pm$0.38     \\
$\gamma$              [\kms]           & 33.178$\pm$0.006  & 34.185$\pm$0.002    & 34.058$\pm$0.003  \\
$\sigma$(O-C)         [\ms]            & 12.74             & 12.28               & 14.43             \\
\hline
\end{tabular}
}

\end{table}

\section{Frequency of hot Jupiters in OCs}

 The  most striking result is that 
 with the star YBP401 we have found three HJs around 66 MS and subgiant stars
 in M67 (53 stars if we only consider single stars). 
 This gives a frequency of HJs of 4.5$^{+4.5}_{-2.5}$\% and 5.6$^{+5.4}_{-2.6}$\%. 
 These results also agree with the HJs frequency (5.5$^{+5.5}_{-2.5}$\%) 
 obtained by a Monte Carlo analysis in our parallel work
 (Brucalassi et al. 2016 sub.).
 Our values are higher than those derived from the RV surveys around FGK stars
 \citep[$1.2\%\pm0.38$ of HJs][]{Wright2012}.
 The comparison is even more striking when considering that the {\em Kepler}
 \footnote{http://kepler.nasa.gov/} statistics of HJs is 
 lower, around 0.4$\%$ \citep{Howard11}.  
 However, the comparison between different samples and between simulations and observations is not trivial. 
 The analysis of the {\em Kepler} and the RV surveys 
 for instance use different selection criteria (radii vs. masses)   
 and different intervals of orbital periods. 
 \citet{Dawson2013} showed that the discrepancy might be partially due
 to the different metallicity of the samples.
 Another effect to take into account is in the definition of the comparison samples.
 RV surveys are performed on pre-selected samples that have been
 corrected for the presence of binaries, 
 while the {\em Kepler} statistics (and most of the simulations) 
 refer to all FGK stars in the Cygnus field, without any previous selection for binaries. 
 For M67 our survey sample was heavily pre-selected with the aim 
 to eliminate all known and suspected binaries in advance.  
 We therefore expect that when we compare our results 
 on the whole sample (3/66 or 4.5$^{+4.5}_{-2.5}$\%) with the {\em Kepler} statistics (0.4$\%$),
 an upper limit of the planet frequency will be provided, while the comparison of the frequency  
 of the single-star sample (3/53 or  5.6$^{+5.4}_{-2.6}$\% ) is expected to compare well 
 with the 1.2$\%$ of the radial velocity surveys because they have gone through a similar selection process.
 Finally, based on a HJ occurrance rate of 1.2\% like for field stars,
 one or two additional HJs may exist in M67 with a non-negligible probability of ~5\%.

 For several years, the lack of detected planets in OCs was in contrast with the field results, 
 but the recent discoveries \citep{Malavolta2016,Brucalassi2014,Meibom2013,Quinn2014,Quinn2012} 
 have completely changed the situation.\\
 These results are difficult to reconcile with the early survey of \citet{Paulson2004}, 
 who did not detect any HJs around 94 G-M stars of the Hyades cluster. 
 Since the extrapolation from  non-detection to non-existence of HJs 
 critically depends on several assumptions such as stellar noise and real measurement errors,
 constraints on the allocated time, number of observations per
 star, sampling and planet mass,
 we cannot state at present whether the discrepancy is real until larger surveys are performed.
 If we were to add the results of the RV surveys 
 in the OCs M67, Hyades, and Praesepe, we would determine 
 a rate of 6 out of 240 HJs per surveyed stars (including some binaries), 
 which is a high percentage when compared to 10 out of 836 HJ
 per surveyed stars in the field sample of \citet{Wright2012}.
 We can conclude that, contrary to early reports, 
 the frequency of HJs discovered in the three OCs 
 subject of recent RV surveys is higher than amongst the field stars. 
 To explain the high frequency of HJs in M67, Hyades, and Praesepe, we may argue that 
 the frequency  of  HJs  depends on stellar metallicity, mass, or on dynamical history, 
 and therefore environment. 
 The dependence  of planet frequency on stellar metallicity is complex: even if established very early 
 \citep{Johnson2010, Udry2007,Fischer2005,Santos2004}, 
 a real correlation seems to be present only for Jupiters 
 around MS stars, while it does not hold for giant planets 
 around evolved stars \citep{Pasquini2007} or for low-mass planets \citep{Mayor2011}. 
 Both Hyades and Praesepe are metal rich 
 \citep{Pace2008,Ferreira2015},
 and this may explain the higher frequency of HJs in these clusters,   
 but this is not the case of M67, which has a well-established solar metallicity and abundance pattern 
 \citep{Randich2005}.  
 The hypothesis that the high frequency of HJs in M67 
 or in OCs in general originates from the higher mass of the host star can be also excluded:
 the stars hosting HJs in M67 all have masses around one solar mass, 
 which is very similar to the masses of the HJ hosts discovered in the field.
 A similar argument holds for the stars hosting HJs in Praesepe and in the Hyades.
 
 Finally,
 environment is left as the most suitable option to explain the HJs excess.  
 It has been suggested that dense birth-environments such as
 stellar clusters can have a significant effect on
 the planet formation process and the resulting orbital properties
 of single planets or planetary systems. Close stellar fly-by or binary
 companions can alter the structure of any planetary system
 and may also trigger subsequent planet-planet scattering
 over very long timescales \citep{Davies2014, Malmberg2011}. 
 This leads to the ejection of some planets, 
 but it also seems to favour the conditions for the formation of HJs \citep{Shara2016}.
 As predicted by such mechanisms, M67 HJs show orbits with non-zero eccentricity, 
 which is also true for the HJ found in the Hyades. 
 The importance of the encounters is primarly determined
 by the local stellar density, the binary fraction, 
 the collisional cross-section of the planetary system, and 
 the timescale on which the planet is exposed to external perturbations.
 \citet{Malmberg2011} produced simulations for a cluster of 700 stars 
 and an initial half-mass radius of 0.38 parsecs,
 showing that a non-negligible number of stars spend long enough as a binary system 
 and also that the majority of the stars is affected by at least one fly-by. 
 M67 has more than 1400 stars at present, it is dominated by a high fraction of binaries 
 \citep{Davenport2010} after loosing at least three quarters 
 of its original stellar mass, and has suffered mass segregation. 
 \citet{Shara2016} have recently completed N-body simulations for a case similar to the one of M67, but
 with only a 10$\%$ of binaries, finding that HJs can be produced in 0.4$\%$ of cluster planetary systems
 when only considering initial fly-by encounters.
 This fraction is smaller than what we find in M67, and
 the influence of other migration mechanisms probably needs to be considered as well to explain our results.
 However, the same authors acknowledged that a higher fraction of binaries
 will strongly enhance the probability of HJ formation and therefore their frequency. 
 Given that the binary fraction of M67 stars is currently very high \citep{Pasquini2012,Mathieu1990} 
 and that models show that it must also have been high at the origin \citep{Hurley2005},
 the high fraction of M67 HJs seems to qualitatively agree with the N-body simulations. 
 The same simulations predict that after 5 Gyr 
 the percentage of stars hosting HJs retained by the cluster
 is substantially higher than the percentage of stars not hosting HJs.
 Two factors can contribute to enhance HJ planet formation: the capability of producing HJs, and the capability 
 of the cluster to retain stars hosting HJs.
 The interaction takes part well within the first Gyr of the cluster lifetime, 
 so that the stars with HJs do not require to still have a stellar companion at the age of M67.
 \citet{Geller2015} reported no evidence for nearby companions at the present epoch.
 Finally, considering that about one of ten HJs produces a transit, 
 we suggest to carefully examine the {\em Kepler}/K2 observations 
 \citep{Howell2014} for any transit. 
 
 \begin{acknowledgements}
 LP acknowledges the Visiting Researcher program of the CNPq Brazilian Agency, at the
 Fed. Univ. of Rio Grande do Norte, Brazil.
 RPS thanks ESO DGDF, the HET project, the PNPS and PNP of INSU - CNRS 
 for allocating the observations.
 MTR received support from PFB06 CATA (CONICYT).
\end{acknowledgements}





\bibliographystyle{aa} 
\bibliography{master} 
 

\Online

%
\begin{table*}
\caption{RV measurements, RV uncertainties, bisector span, and ratio of the H$_{\alpha}$ core 
  with respect to the continuum \citep[see][]{Pasquini1991} for YBP401. 
  All the RV data points are corrected to the zero point of HARPS.}             
\label{table_YBP401}      
\centering          
\begin{tabular}{lrrrrl }     
\hline\hline        
  BJD & RV & $\sigma_{obs}$ & BIS span& H$_{\alpha}$ratio& instrument\\
  (-2450000) & (\kms) & (\kms)& (\kms) &  & \\
\hline                    
  4491.3462     &  33.2165     &  0.0089   &$-$0.0013333  & 0.03133    & SOPHIE    \\%
  4855.6494     &  33.1522     &  0.0110   &$-$0.0241850  & 0.02735    & HARPS    \\%
  4859.5234     &  33.1978     &  0.0089   &   0.0150000  & 0.02843    & SOPHIE    \\%
  4861.8052     &  33.1559     &  0.0110   &   0.0573989  & 0.02435    & HARPS     \\%
  5190.8584     &  33.1793     &  0.0080   &$-$0.0384314  & 0.02154    & HARPS     \\      
  5627.7314     &  33.2302     &  0.0270   &   0.0061686  & 0.02206    & HARPS     \\      
  5946.8267     &  33.1927     &  0.0180   &   0.0812266  & 0.02624    & HARPS       \\    
  5967.6030     &  33.1589     &  0.0170   &$-$0.0273837  & 0.02377    & HARPS      \\     
  5978.5732     &  33.2331     &  0.0170   &   0.0187173  & 0.02635    & HARPS      \\     
  5985.5850     &  33.1612     &  0.0090   &$-$0.0098333  & 0.02806    & SOPHIE       \\   
  6245.8511     &  33.1369     &  0.0150   &   0.0556374  & 0.02424    & HARPS       \\    
  6269.7925     &  33.1744     &  0.0170   &   0.0426396  & 0.02201    & HARPS       \\    
  6302.7832     &  33.1737     &  0.0220   &   0.0607550  & 0.02471    & HARPS       \\%
  6308.7578     &  33.1658     &  0.0110   &   0.0069258  & 0.02561    & HARPS        \\   
  6323.7534     &  33.1513     &  0.0170   &   0.0179823  & 0.02399    & HARPS        \\   
  6333.7314     &  33.2038     &  0.0140   &   0.0456952  & 0.02144    & HARPS       \\    
  6682.6787     &  33.1879     &  0.0090   &   0.0310414  & 0.02459    & HARPS        \\   
  6694.5210     &  33.1934     &  0.0124   &   0.0562494  &   -        & HARPS-N        \\ 
  6697.4912     &  33.2130     &  0.0092   &   0.0208414  &   -        & HARPS-N      \\   
  6715.7002     &  33.1711     &  0.0110   &   0.0338823  & 0.02443    & HARPS        \\%
  6719.6357     &  33.1821     &  0.0140   &   0.0273266  & 0.02430    & HARPS        \\%
  6720.4028     &  33.1154     &  0.0145   &$-$0.0233333  & 0.02147    & SOPHIE          \\
  6721.5381     &  33.1395     &  0.0129   &   0.0215000  & 0.03277    & SOPHIE          \\
  6977.8101     &  33.1514     &  0.0150   &$-$0.0060115  & 0.02535    & HARPS         \\ 
  6978.8369     &  33.1645     &  0.0160   &   0.0337306  & 0.02274    & HARPS          \\ 
  6983.8188     &  33.2158     &  0.0120   &   0.0234168  & 0.02289    & HARPS         \\ 
\hline                                                                         
\end{tabular}                                                              
\end{table*}
 \begin{table*}
\caption{Orbital parameters of the planetary candidates using
 a simple MCMC analysis to fit Keplerian orbits to the RV data.
 We considered as free parameters the orbital period $P$, the time of transit T$_{c}$, the radial
 velocity semi-amplitude $K$, the centre-of-mass velocity $\gamma$, and
 the orthogonal quantities $\sqrt{e}\cos \omega$ and $\sqrt{e}\sin \omega$, where
 $e$ is the eccentricity and $\omega$ is the argument of periastron.
 We quote the mode of the resulting parameter distributions as the final value and
 the $68.3\%$ interval with equal probability density at the $\pm1\sigma$ bound to derive the uncertainty.
 \ensuremath{T}: time at periastron passage,
 \ensuremath{K}: semi-amplitude of the RV curve, 
 \ensuremath{m\sin{i}}: planetary minimum mass.}
\label{PlanetParamMCMC}
\centering
\begin{tabular}{lrrr}
\hline
\textbf{Parameters} &YBP401&YBP1194&YBP1514\\
\hline
\ensuremath{P}        $[\mathrm{days}]$& 4.0873$^{+0.0003}_{-0.0002}$    & 6.959$^{+0.001}_{-0.001}$       & 5.1189$^{+0.0008}_{-0.0007}$   \\
\ensuremath{T}        $[\mathrm{JD}]$  & 2455974.23$^{+0.49}_{-0.49}$    & 2455290.0$^{+0.4}_{-0.3}$       & 245986.34$^{+0.28}_{-0.20}$ \\
\ensuremath{e}                         & 0.141$^{+0.112}_{-0.113}$       & 0.294$^{+0.077}_{-0.056}$       & 0.332$^{+0.133}_{-0.127}$   \\
$\omega$              $[\mathrm{deg}]$ & -31.69$^{+43.0}_{-54.0}$        & 99.14$^{+16.0}_{-16.0}$         & -34.76$^{+17.35}_{-14.26}$  \\
$\sqrt{e}\sin \omega$                  & -0.197$^{+0.152}_{-0.166}$      & 0.535$^{+0.061}_{-0.064}$       & -0.329$^{+0.134}_{-0.136}$                 \\
$\sqrt{e}\cos \omega$                  &  0.319$^{+0.104}_{-0.242}$      &-0.086$^{+0.186}_{-0.171}$       & 0.474$^{+0.092}_{-0.086}$                 \\
\ensuremath{K}        [\ms]            &  48.911$^{+5.0}_{-6.0}$         & 35.607$^{+2.2}_{-4.0}$          & 50.97$^{+3.0}_{-3.0}$  \\
\ensuremath{m\sin{i}} [\Mj]            &  0.41$^{+0.06}_{-0.04}$         & 0.32$^{+0.3}_{-0.2}$            & 0.42$^{+0.03}_{-0.02}$     \\
$\gamma$              [\kms]           &  33.172$^{+0.003}_{-0.004}$     & 34.185$^{+0.002}_{-0.002}$      & 34.058$^{+0.003}_{-0.003}$  \\
\hline
\end{tabular}

\end{table*}

\end{document}